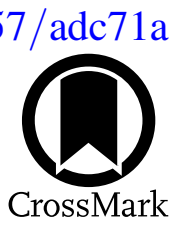

# Asteroseismic Analysis of a Red Giant KIC 9145955 by Including the Small-scale Magnetic Fields in the Atmosphere

Yuetong Wang[1,2,3], Guifang Lin[1,3], Yan Li[1,2,3,4], Tao Wu[1,3,4], and Yaguang Li[5]

[1] Yunnan Observatories, Chinese Academy of Sciences, Kunming 650216, People's Republic of China; wangyuetong@ynao.ac.cn
[2] University of Chinese Academy of Sciences, Beijing 100049, People's Republic of China
[3] International Center of Supernovae at the Yunnan Key Laboratory, Kunming 650216, People's Republic of China
[4] Center for Astronomical Mega-Science, Chinese Academy of Sciences, Beijing 100012, People's Republic of China
[5] Institute for Astronomy, University of Hawai'i, 2680 Woodlawn Drive, Honolulu, HI 96822, USA

## Abstract

Recent convincing evidence is found within asteroseismology that suggests the magnetic fields exist in three red giants. Research on small-scale magnetic fields in the Sun and HD 49385 has shown that they have a certain corrective effect on the systematic discrepancies between observed and theoretical frequencies. Here we apply a similar method applied for the Sun to a red giant, KIC 9145955, to explore the impact of small-scale magnetic fields in the photosphere on its frequencies. We find that the calculated frequencies of our best-fit model, which simulates the effect of the magnetic fields by artificially modifying the Eddington $T-\tau$ relation, perfectly match those of the observed $l = 0$, 1, and 2 modes, indicating the existence of small-scale magnetic fields with an upper strength limit of 65 G and concentrating at a height 13,100 km in the photosphere. Based on the best-fit model, we revise the stellar parameters of KIC 9145955 as $M = 1.23 \pm 0.04\,M_\odot$, $R = 5.57 \pm 0.06\,R_\odot$, $L = 19.85 \pm 0.5\,L_\odot$, Age $= 3.83 \pm 0.5$ Gyr, $M_{\rm He} = 0.2108 \pm 0.0005\,M_\odot$, and $R_{\rm He} = 0.0306 \pm 0.0001\,R_\odot$.

## 1. Introduction

Magnetic fields are widely present in stars and are an essential component in the stellar structure and evolution. They are the cornerstone of all kinds of atmospheric activities on the Sun and stars, including photospheric spots and flares, and coronal mass ejections (CMEs) from the chromosphere and corona. The reconnection of the magnetic fields plays a crucial role in the initiation, acceleration, and final energy release of CMEs (J. Lin & T. G. Forbes 2000; J. Lin et al. 2015; B. Schmieder et al. 2024). The magnetic fields influence the spin-down of the stars through modulating the stellar winds (M. J. Aschwanden 2004; E. Priest 2014; A. K. Srivastava et al. 2024). Moreover, the magnetic fields play a pivotal role in the stellar structure and evolution. For instance, the presence of magnetic braking (MB) can impact the age determination of massive stars, and surface fossil magnetic fields can significantly reduce the stellar mass-loss rate (A. David-Uraz et al. 2021; Z. Keszthelyi et al. 2021, 2022, 2023, 2024). M. Echeveste et al. (2024) conducted a study on the evolution of low-mass X-ray binaries under different MB prescriptions, which shows the impact of MB on angular momentum loss and the ability to maintain stable mass transfer. They also investigated how these factors influence the orbital period evolution and the eventual state of the system. Therefore, it is crucial to discover the presence of magnetic fields in stars and determine their related properties.

Through three-dimensional numerical simulations, L. T. Lehmann et al. (2018) studied the impact of stellar parameters on magnetic field topology and different manifestations in small-scale and large-scale magnetic fields. The Zeeman–Doppler Imaging technique (M. Semel 1989; J.-F. Donati & S. F. Brown 1997; J.-F. Donati et al. 2006) can only observe the large-scale magnetic field topology for remote stars, and its resolution depends on many factors, such as the star's rotational velocity (C. Johnstone et al. 2010; D. Arzoumanian et al. 2011; P. Lang et al. 2014). Based on the studies of $\gamma$ Doradus ($\gamma$ Dor) stars and slowly pulsating B-type stars, F. Lignières et al. (2024) explored the impact of magnetic fields on the frequency shifts of high-order gravity modes and Rossby modes. With the aid of asteroseismology, G. Li et al. (2022) analyzed the mixed modes of three red giant stars and confirmed the frequency shifts induced by magnetic fields. By comparing observational data with theoretical models, they found the presence of magnetic fields ranging from 30 to 100 kG near the hydrogen-burning shell of the three stars. Y. Li et al. (2021) used the frequency inconsistency to derive that the Sun has small-scale magnetic fields of 90 G at a height of 630 km above the photosphere, consistent with observational and numerical simulation results (J. Trujillo Bueno et al. 2004; B. W. Lites et al. 2008; O. Steiner et al. 2008; M. Rempel 2014). Y. Wang et al. (2024) conducted a detailed study on the frequencies of solar-like modes of HD 49385 with a method analogous to Y. Li et al.'s (2021) approach, and deduced the presence of small-scale magnetic fields with a strength of 80 G at a height of 1850 km in its atmosphere.

J. Christensen-Dalsgaard et al. (1988) and W. A. Dziembowski et al. (1988) identified a systematic discrepancy between the theoretical frequencies of the standard solar model and the observed frequencies. This discrepancy is independent of the spherical degree $l$ and increases with frequency. This phenomenon is known as the solar surface effect (J. Christensen-Dalsgaard et al. 1988; W. A. Dziembowski et al. 1988). It is generally believed that the solar surface effect arises from the highly complex structure of the solar surface, including turbulent pressure, nonadiabatic effects, rapid variations in thermodynamic quantities, and atmospheric magnetic fields (T. M. Brown 1984; G. Houdek et al. 2017; Y. Li et al. 2021). To mitigate the impact of surface effects, numerous correction formulas have been proposed. For instance, H. Kjeldsen et al. (2008) introduced a power-law correction formula to adjust theoretical frequencies, which has been widely adopted for other stars (e.g., J. Christensen-Dalsgaard et al. 2010; Y. K. Tang & N. Gai 2011; S. Deheuvels et al. 2012; S. Mathur et al. 2012). Based on D. O. Gough (1990) asymptotic analysis, W. H. Ball & L. Gizon (2014, 2017) developed correction formulas that yielded excellent results for radial modes on the Sun and red-giant-branch stars. Moreover, modeling stellar surfaces with three-dimensional hydrodynamical simulations has significantly improved stellar models (e.g., H. G. Ludwig et al. 2009; B. Beeck et al. 2013; R. Trampedach et al. 2013). T. Sonoi et al. (2015, 2019b) demonstrated the influence of thermodynamic effects on surface corrections using three-dimensional numerical simulations of the atmosphere, achieving notable results. Y. Li et al. (2021) investigated the impact of small-scale magnetic fields in the solar photosphere on the oscillation frequencies of solar $p$-modes. They found that solar $p$-mode oscillations are reflected at the magnetic-arch splicing layer, which helps to improve the systematic discrepancies between observed and theoretical frequencies.

In the present work, we aim to extend the above method of exploring the small-scale magnetic fields to an important class of solar-like pulsators: the red giants. Therefore, we selected KIC 9145955, a thoroughly studied star, as the target of our research. The extensive previous research on this star has provided an excellent foundation for our work. KIC 9145955 is a bright star located on the red giant branch, with observational data obtained from Kepler (W. Borucki et al. 2008). Numerous studies have been focused on it, since it has abundant high-precision observational data and shows no complexity of power density spectra caused by the rotational splitting (T. R. Bedding et al. 2011; B. Mosser et al. 2012; S. Hekker & A. Mazumdar 2014; E. Corsaro et al. 2015; A. Datta et al. 2015). X. Zhang et al. (2018) conducted a detailed asteroseismic analysis for KIC 9145955, with precisely determined fundamental stellar parameters and the mass of its helium core. However, a significant discrepancy remains between the observed frequencies and those of the best-fitting theoretical models for $l = 0$ and 2 modes, without applying the empirical correction formulas for the so-called surface effect (H. Kjeldsen et al. 2008; W. H. Ball & L. Gizon 2014, 2017). We shall modify the standard atmospheric $T{-}\tau$ relation in our models to simulate the effect of small-scale magnetic fields in the stellar photosphere, and conduct a detailed asteroseismic analysis for KIC 9145955 to examine the impact of small-scale magnetic fields on the stellar oscillation frequencies. This approach can provide information on the small-scale magnetic fields in the red giants. However, both thermodynamic effects (nonadiabaticity and compressible turbulence) and magnetic fields contribute to surface effects. Due to current computational limitations, it is not feasible to incorporate adjustable magnetic field parameters into atmosphere models with 3D hydrodynamical simulations. As a result, calculations can only be conducted based on the classical Eddington gray atmosphere model. As thermodynamic contributions to the correction of stellar models were not considered, we are likely to have overestimated the role of magnetic fields. Therefore, the values obtained represent the upper limits of small-scale magnetic fields. It is hoped that future studies will simultaneously account for both thermodynamic effects and magnetic fields to further refine theoretical stellar models.

In this paper, we use the oscillation frequencies extracted and validated by X. Zhang et al. (2018) and the small-scale magnetic fields and boundary conditions proposed by Y. Li et al. (2021) for asteroseismic calculations. Section 2 details the input physical parameters for asteroseismic models, modifications to the magnetic fields and boundary conditions, and the criteria for model selection. Section 3 presents the process and model fitting results, followed by in-depth discussions in Section 4. Finally, Section 5 provides a summary of our work.

## 2. Asteroseismic Models

### 2.1. Input Physics

We employ the Modules for Experiments in Stellar Astrophysics (MESA; version 10398; B. Paxton et al. 2011, 2013, 2015, 2018, 2019; A. S. Jermyn et al. 2023) program to construct our stellar models in this study. We utilize the ADIPLS module within the MESA program to compute stellar pulsation frequencies. The ADIPLS module is an adiabatic pulsation program released by J. Christensen-Dalsgaard (2008).

The fundamental input parameters we have adopted in this study are as follows. The equation of state used by us is the version published by F. J. Rogers & A. Nayfonov (2002). We employ the opacity tables published by C. A. Iglesias & F. J. Rogers (1996) for the high-temperature region, and the opacity tables released by J. W. Ferguson et al. (2005) for the low-temperature region. The stellar atmosphere model adopted by us is the Eddington gray atmosphere model. Specific details regarding the inclusion of the magnetic field and modifications to the boundary conditions will be elaborated in the following section. Stellar convection is treated using the mixing-length theory (E. Böhm-Vitense 1958). For KIC 9145955, we do not consider rotation, diffusion, and convective overshooting in the models.

### 2.2. Adding Magnetic Field and Modifying Mechanical Surface Boundary Condition

We employ the Eddington gray atmosphere model (C. Aerts et al. 2010) and consider the magnetic field effects in the photosphere. To construct the internal and surface structure of KIC 9145955, we build upon the work of Y. Li et al. (2021), which examines the impact of small-scale magnetic fields in the photosphere on solar oscillation frequencies. The effect of the magnetic fields is approximately introduced into the stellar photosphere with the aid of an additional term in the Eddington $T{-}\tau$ relation, which simulates the effect of the magnetic pressure. The atmospheric $T{-}\tau$ relation can be uniformly expressed as

$$T^4 = \frac{3}{4} T_{\rm eff}^4 [\tau + q(\tau)], \tag{1}$$

where $T$ represents the temperature, $T_{\rm eff}$ is the effective temperature of the star, $\tau$ denotes the optical depth from the stellar surface, and $q(\tau)$ is a Hopf-like function. As noted by T. Sonoi et al. (2019a), different $T{-}\tau$ relations significantly affect the entropy profile in the solar atmosphere. Their Equation (3) provides a more complex and accurate description than the Eddington $T{-}\tau$ relation, which is based on more physical processes and observational data. The VAL-C $T-\tau$ relation better reproduces the lowest entropy in the three-dimensional model of the solar photosphere. However, the physical structure of KIC 9145955's atmosphere is very much different from that of the Sun due to quite different effective temperatures and gravitational accelerations. Therefore, we simply assume the Eddington gray atmosphere with $q = 2/3$. Based on Y. Li et al. (2021), we introduce an additional term into the Eddington $T-\tau$ relation, which phenomenologically simulates the effect of the magnetic fields in the photosphere of KIC 9145955. Then, the Hopf-like function $q(\tau)$ that we have adopted is

$$q(\tau) = \frac{2}{3} + a\exp(-b\sqrt{\tau}), \tag{2}$$

where the two parameters we have introduced, $a$ and $b$, represent, respectively, the strength of the magnetic fields and their location in the atmosphere. It can be noticed that the second term on the right-hand side of Equation (2) causes the temperature to rise if the optical depth falls below a critical level, thereby resulting in an increase of the radiation pressure. Such an artificial increase of the pressure is regarded to simulate the effect of the magnetic fields in the upper atmosphere. By adjusting the values of $a$ and $b$, we can better reproduce the observed oscillation frequencies on KIC 9145955 and finally infer the properties of the small-scale magnetic fields in its upper atmosphere.

The surface mechanical boundary condition is a crucial factor in the propagation and reflection of stellar oscillations in the magnetized upper atmosphere, as already discussed by Y. Li et al. (2021). The results indicate that employing Eulerian perturbation boundary conditions expands the propagation region of acoustic waves, thereby increasing the acoustic wave travel time and leading to a decrease in frequency (for detailed information, see Figures 2 and 3 in Y. Li et al. 2021). From a physical perspective, when magnetic pressure increases rapidly, the gas pressure decreases, resulting in a rapid decline in density and a corresponding increase in sound speed. Therefore, stellar oscillations should be reflected at the magnetic-arch splicing layer. Thus, the traditional boundary condition for linear adiabatic oscillations (W. Unno et al. 1989; J. Christensen-Dalsgaard 2008),

$$\delta P = P' - \rho g \xi_r = 0, \tag{3}$$

is no longer valid. Based on the discussions of Y. Li et al. (2021), we adopt the surface mechanical boundary condition as

$$P' = 0, \tag{4}$$

where $\xi_r$ is the radial displacement, $P'$ is the Eulerian perturbation of pressure, $g$ is the gravitational acceleration, and $\rho$ is the density. The comparison of pulsation models using the aforementioned boundary conditions is shown in the Appendix.

### 2.3. Model Fittings

In our work, the initial helium abundance $Y$ is assumed to be related to the initial metal abundance $Z$ by (A. Dotter et al. 2008)

$$Y = Y_0 + \frac{dY}{dZ}Z. \tag{5}$$

Here, we assume that the value of $Y_0 = 0.245$ and $dY/dZ = 1.54$ (B. Thompson et al. 2014). We use the following formula of $[Z/X]$ to obtain the value of $Z$,

$$[Z/X] = \log\left(\frac{Z}{X}\right) - \log\left(\frac{Z}{X}\right)_\odot. \tag{6}$$

Here, we adopt that $(Z/X)_\odot = 0.0245$ (N. Grevesse & A. Noels 1993) and the $[Z/X]$ value is obtained from F. Pérez Hernández et al. (2016) for KIC 9145955.

In our calculations, each star with the given values of mass and metallicity evolves from the pre-main-sequence stage to the red giant stage. We also use luminosity as a constraint for frequency calculations based on the work of X. Zhang et al. (2018). When the star satisfies the luminosity condition $15L_\odot < L < 23L_\odot$ during the red giant branch (RGB) stage, we calculate the frequencies for $l = 0$, 1, and 2 oscillation modes. Based on the results of X. Zhang et al. (2018; Figure 7 in their paper) and our preliminary results, we further shorten the time steps of the evolution models in the range $3.03 < \log g < 3.042$ to ensure that the best model is accurately identified.

In order to evaluate the goodness of frequency fitting between the observed data and the theoretical model, we define two quantities:

$$\chi^2_{\rm all} = \frac{1}{N_{\rm all}}\sum_{i=1}^{N_{\rm all}}(\nu^{\rm obs}_{i({\rm all})} - \nu^{\rm mod}_{i({\rm all})})^2 \tag{7}$$

and

$$\chi^2_{l=0} = \frac{1}{N_{l=0}}\sum_{i=1}^{N_{l=0}}(\nu^{\rm obs}_{i(l=0)} - \nu^{\rm mod}_{i(l=0)})^2\,, \tag{8}$$

where $\nu_i^{\rm obs}$ represents the observed frequencies and $\nu_i^{\rm mod}$ represents the calculate ones, and $N$ is the number of the observed frequencies. Equation (7) represents the matching results for all modes, and Equation (8) represents the matching results for $l = 0$ modes only. Here, we do not incorporate the errors of the observed frequencies into the goodness-of-fit functions. As X. Zhang et al. (2018) discussed, the stellar rotation causes nonradial modes to split into multiple frequencies, but the rotational splitting phenomenon was not observed, which leads to uncertainties in frequency determination. Therefore, we believe that it is reasonable not to include these errors in the fit.

## 3. Fitting Results of KIC 9145955

Considering the broad parameter space comprising four parameters (i.e., $a$, $b$, $M$, and $\alpha$), a comprehensive grid search would demand computational resources far beyond our available capacity. Following the methodology employed by our previous work of HD 49385 (Y. Wang et al. 2024) and G. Lin et al. (2024) in that of KIC 8006161, they initially scanned the magnetic field parameters $a$ and $b$ under a fixed set of $M$ and $\alpha$ values. After identifying the optimal $a$ and $b$, they then fixed these magnetic field parameters and proceeded to scan $M$ and $\alpha$, ultimately deriving the corresponding stellar parameters. Considering that HD 49385 is a star near the main sequence, variations in $M$ have minimal impact on its radius, implying that the height of its atmospheric layers remains relatively unaffected. Consequently, the corresponding small-scale magnetic field parameters are also insignificantly influenced. In contrast, for KIC 9145955, applying the same approach may introduce some discrepancies in the derived magnetic field parameters.

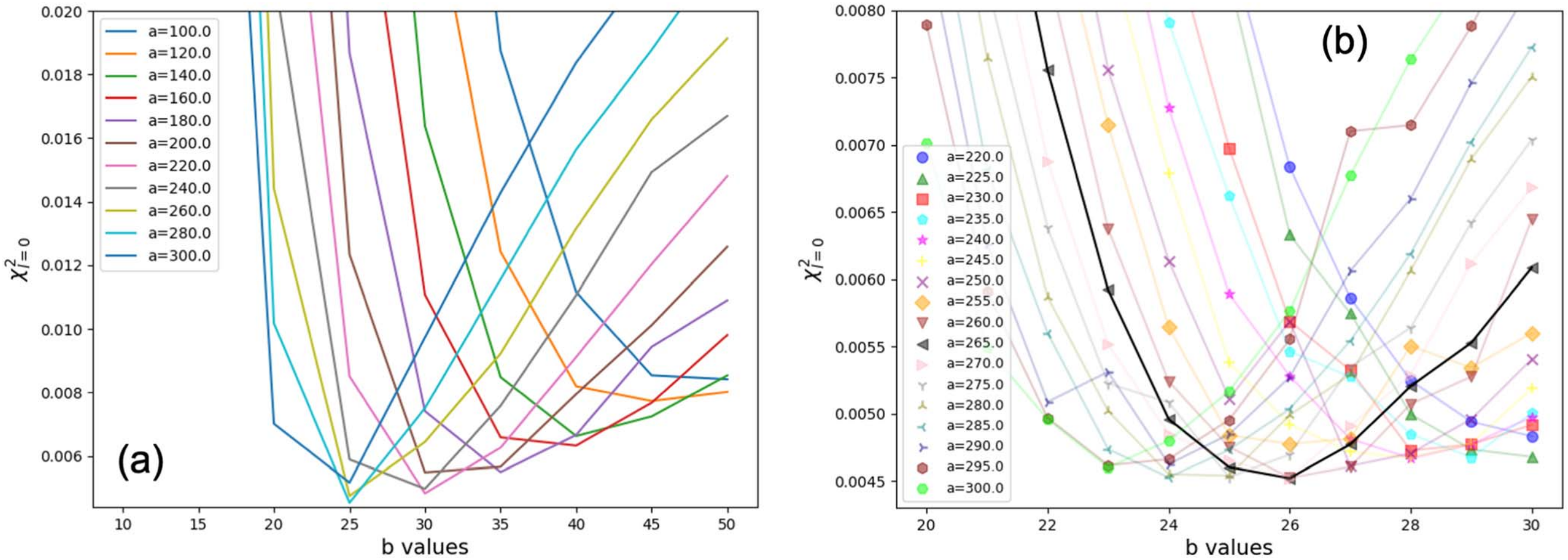

**Figure 1.** (a) Performance of the best-fit models on the coarse grid of the stellar parameters. (b) The performance of the magnetic field parameters in the second refined grid. The $x$-axis represents the $b$ value, and the $y$-axis represents the fitting goodness $\chi^2_{l=0}$. Each line on the graph corresponds to the fitting results with a specific value of $a$. Specifically, all stellar models shown here are calculated with $M = 1.22M_\odot$ and the mixing-length parameter $\alpha = 2.0$.

To establish a more accurate theoretical model, we initially performed a coarse scan over the parameters $M$, $a$, and $b$ to delineate the relationship between $M$ and the magnetic field (Figures 1 and 2). As shown in Figure 2, in the subsequent refined scan over $a$ and $b$ within a narrow range, variations in $M$ have little impact on the magnetic field parameters. Consequently, we fixed $M$ at $1.22M_\odot$ and conducted a refined, narrow-range scan of $a$ and $b$ to derive the magnetic field parameters.

With the magnetic field parameters $a$ and $b$ thus determined, we then performed a grid scan over $M$ and $\alpha$ to obtain our final model results. The detailed scanning process and grid settings are as follows.

First, we calculate stellar models in a grid of stellar parameters consisting of $M$ from $1.15M_\odot$ to $1.29M_\odot$ with a step of 0.01, $a$ from 100 to 300 with a step of 20, and $b$ from 10 to 50 with a step of 5. By matching the calculated frequencies with the observed ones for $l = 0$ modes, we determine the optimal combination of $M$, $a$, and $b$ as shown in Figure 1(a). This coarse grid with larger steps of the stellar parameters and including only $l = 0$ modes in the frequency match is specifically designed to search for the best-fit combination of the magnetic field parameters $a$, $b$, and the stellar mass $M$. It can be seen clearly that the best-fit model appears at the combination of the magnetic field parameters, $a = 280$ and $b = 25$. The red line in Figure 2 illustrates the fitting performance of the stellar models with $a = 280$ and $b = 25$, but under different values of $M$. The value of $M = 1.22M_\odot$ corresponds to the smallest value of $\chi^2_{l=0}$. So we take $M = 1.22M_\odot$ as the next starting point and conduct a search for $a$ and $b$ in a more refined grid consisting of $a = 220$ to 300 with a step of 5 and $b = 20$ to 30 with a step of 1, as shown in Figure 1(b). Ultimately, we obtain the best-fit magnetic field parameters: $a = 265$ and $b = 26$.

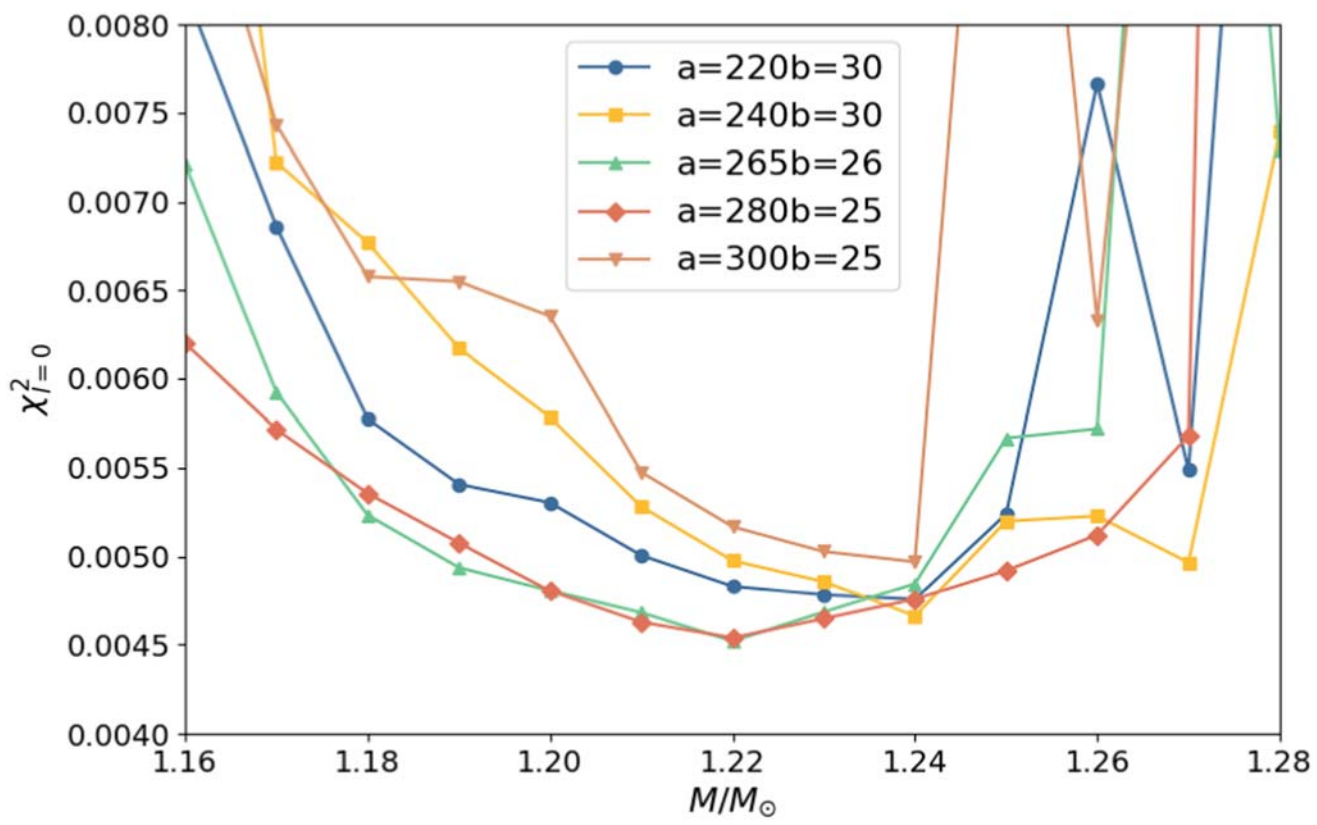

**Figure 2.** Performance of the best-fit models under five sets of magnetic field parameters with varying values of $M$ on the coarse grid of stellar parameters.

Second, we search for the best-fit model of KIC 9145955 with the parameters of the magnetic fields that have just been determined. In order to restrict the best-fit models with the spectroscopic observations, we adopt the observed [Fe/H] value of KIC 9145955 as −0.34 (Y. Takeda et al. 2016). Using Equation (6), we calculate that the corresponding metallicity is $Z \approx 0.0066$. Figure 2 illustrates the variation of $\chi^2_{l=0}$ with $M$ under five sets of magnetic field parameters. $\chi^2_{l=0}$ converges within the range of $1.22M_\odot$ to $1.24M_\odot$, while the asteroseismic result of X. Zhang et al. (2018) is $1.24M_\odot$. We have also preliminarily tested the sensitivity of the fitting performance with respect to the mixing-length parameter $\alpha$ within the range of 1.6–2.2. Based on all of the above considerations, we search for the best-fit model in a concise grid of the stellar parameters, which consists of $M$ from $1.15M_\odot$ to $1.31M_\odot$ with a step of $0.01M_\odot$ and $\alpha$ from 1.95 to 2.05 with a step of 0.01, and all stellar models are computed with a fixed value of $Z = 0.006$.

Based on the grid of stellar parameters discussed above, our theoretical models consist of a total of 187 evolutionary sequences. We use Equation (7) to find the smallest $\chi^2_{\rm all}$ for each evolutionary track, including all of the observed oscillation modes with $l = 0$, 1, and 2. The distribution of $\chi^2_{\rm all}$ with the parameters is shown in Figure 3. As seen in Figure 3(a), $M$ exhibits a good convergence, allowing us to constrain its value to between $1.22M_\odot$ and $1.26M_\odot$. Deviations from this range result in a significant jump in $\chi^2_{\rm all}$. In contrast, $\alpha$ does not display the same level of convergence as mass, as shown in Figure 3(b). This result is consistent with X. Zhang et al. (2018), who also observed that candidate models with varying mixing-length parameters yield very similar $\chi^2_{\rm all}$ values, leading to considerable uncertainty. This may be attributed to red giants having thicker convective envelopes, while $\alpha$ only influences the outermost layers. Figure 4(a) shows the relationship between $\chi^2_{\rm all}$ and $\log g$. It can be seen that the models with $\chi^2_{\rm all} < 0.02$ have their corresponding $\log g$ values concentrated between 3.035 and 3.04. Figure 4(b) demonstrates the relationship between $\chi^2_{\rm all}$ and the stellar helium core mass. Consequently, the helium core mass of KIC 9145955 converges within $M_{\rm He} = 0.2108 \pm 0.0005(M_\odot)$. This is consistent with the results of X. Zhang et al. (2018).

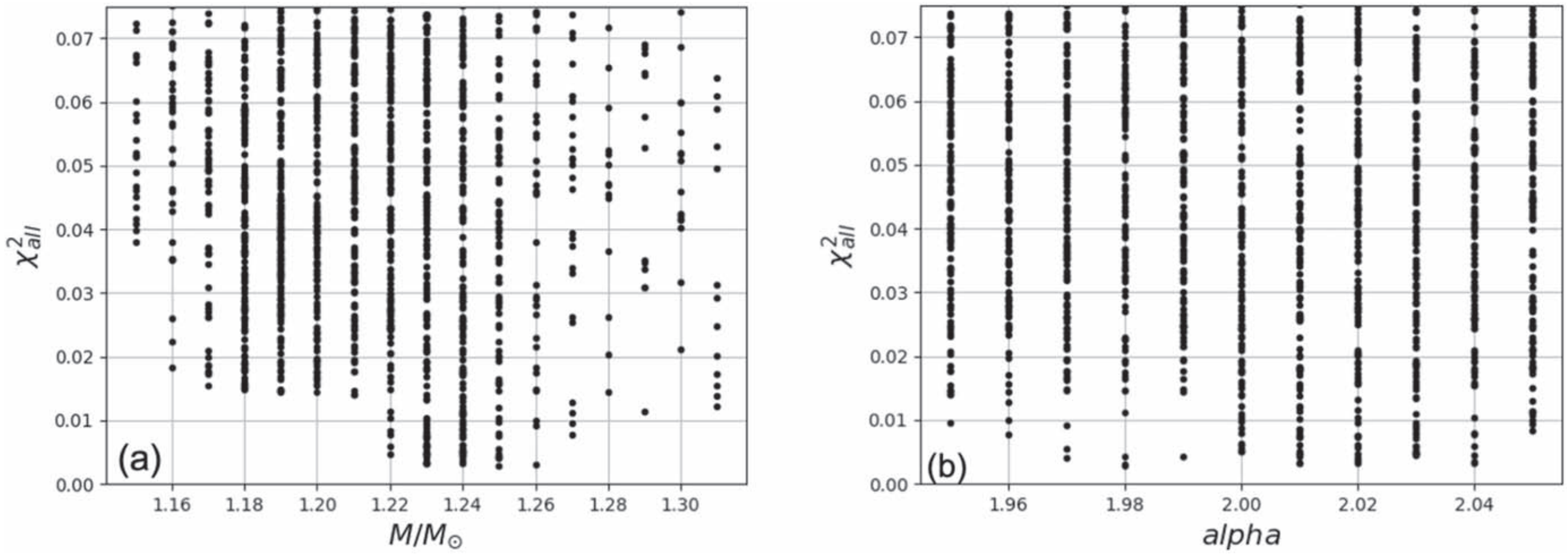

**Figure 3.** The relationship between $\chi^2_{\rm all}$ and the variations in $M$ and $\alpha$, when fixing the magnetic parameters as $a = 265$ and $b = 26$.

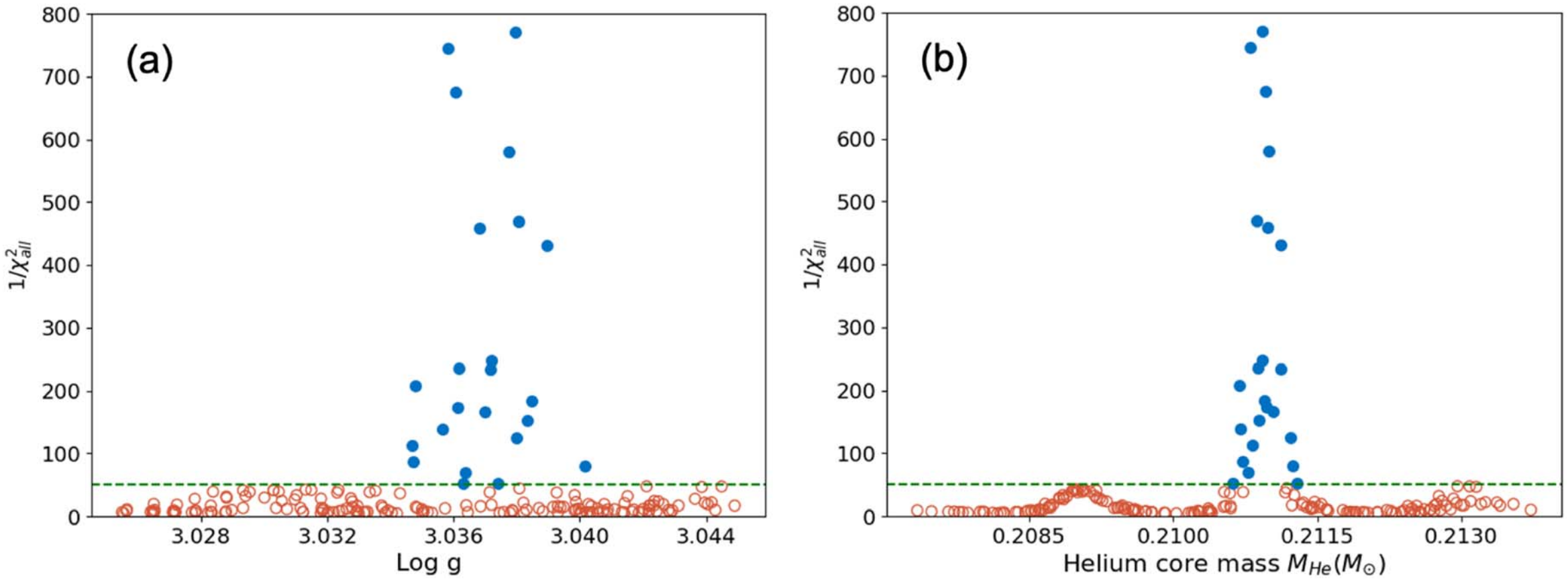

**Figure 4.** (a) $1/\chi^2_{\rm all}$ as a function of $\log g$ for the best-fit models of KIC 9145955. (b) $1/\chi^2_{\rm all}$ as a function of the mass of the helium core for the best-fit models of KIC 9145955. Each circle represents a model with the minimal value of $\chi^2_{\rm all}$ on each evolutionary track. Solid circles represent the candidate models with $\chi^2_{\rm all} < 0.02$, as shown in Table 1.

We select our candidate models based on the criterion $\chi^2_{\rm all} < 0.02$. Table 1 provides the detailed parameters of these models. Figure 5 depicts the evolutionary trajectories of Model E and Model Q. Although Model Q exhibits a smaller $\chi^2_{\rm all}$ than Model E, its $1/\chi^2_{l=p}$ and $\chi^2_{\rm all}$ do not converge at the same point. Consequently, given the close correspondence between the $\chi^2_{\rm all}$ values of Model E and Model Q, we have selected Model E as our best-fit model. Since the fitting accuracy of our selected model ($\chi^2_{\rm all} < 0.02$) is essentially within the range of observational errors (most observational errors fall between 0.1 and 0.2 $\mu$Hz), we provide only an approximate range for parameter deviations. We describe the values of the stellar parameter in the form $x \pm dx$, where $x$ is the value of the best-fit model and $dx$ represents the maximum deviation between the candidate models and the best-fit model. The stellar parameters thus obtained are shown in Table 2, along with a comparison with observational results and other model studies.

Figure 6 shows the differences between the observed frequencies of KIC 9145955 and the theoretical frequencies from our best model, Model E. It can be found that the frequency differences for all observed modes are within 0.1 $\mu$Hz. The Échelle diagram of the observed and calculated frequencies also shows very good agreement between the observations and theoretical models. This feature demonstrates that our theoretical model is in excellent agreement with the observational data.

## 4. Results and Discussions

In the previous section, we select Model E as our best-fit model. From the temperature distribution equations (Equations (1) and (2)), it is evident that the temperature diminishes with decreasing optical depth, reaching a minimum at the stellar

**Table 1**
Values of the Stellar Parameters from Our Candidate Models with $\chi^2_{\rm all} < 0.02$

| Model | $\alpha$ | Initial $M$ ($M_\odot$) | Initial $Z$ | $T_{\rm eff}$ (K) | $\log g$ (dex) | $R$ ($R_\odot$) | $L$ ($L_\odot$) | $\chi^2_{\rm all}$ | $M_{\rm He}$ ($M_\odot$) | $R_{\rm He}$ ($R_\odot$) | $\tau_0$ (day) | Age (Gyr) |
|---|---|---|---|---|---|---|---|---|---|---|---|---|
| A | 2.03 | 1.22 | 0.006 | 5174 | 3.035 | 5.56 | 19.88 | 0.0048 | 0.2107 | 0.0307 | 0.4728 | 3.94 |
| B | 2.04 | 1.22 | 0.006 | 5180 | 3.035 | 5.56 | 19.96 | 0.0116 | 0.2107 | 0.0307 | 0.4728 | 3.94 |
| C | 2.05 | 1.22 | 0.006 | 5185 | 3.035 | 5.56 | 20.05 | 0.0089 | 0.2108 | 0.0307 | 0.4729 | 3.94 |
| D | 2.00 | 1.23 | 0.006 | 5161 | 3.036 | 5.57 | 19.80 | 0.0073 | 0.2107 | 0.0307 | 0.4730 | 3.83 |
| E | 2.01 | 1.23 | 0.006 | 5167 | 3.036 | 5.57 | 19.87 | 0.0013 | 0.2108 | 0.0306 | 0.4729 | 3.83 |
| F | 2.02 | 1.23 | 0.006 | 5172 | 3.036 | 5.57 | 19.94 | 0.0042 | 0.2109 | 0.0306 | 0.4726 | 3.83 |
| G | 2.03 | 1.23 | 0.006 | 5177 | 3.036 | 5.57 | 20.02 | 0.0058 | 0.2110 | 0.0307 | 0.4727 | 3.83 |
| H | 2.04 | 1.23 | 0.006 | 5183 | 3.036 | 5.57 | 20.11 | 0.0015 | 0.2110 | 0.0307 | 0.4727 | 3.83 |
| I | 1.97 | 1.24 | 0.006 | 5148 | 3.036 | 5.59 | 19.72 | 0.0194 | 0.2106 | 0.0306 | 0.4734 | 3.72 |
| J | 1.99 | 1.24 | 0.006 | 5159 | 3.036 | 5.59 | 19.89 | 0.0144 | 0.2108 | 0.0307 | 0.4734 | 3.72 |
| K | 2.00 | 1.24 | 0.006 | 5164 | 3.037 | 5.59 | 19.93 | 0.0040 | 0.2109 | 0.0307 | 0.4727 | 3.72 |
| L | 2.01 | 1.24 | 0.006 | 5169 | 3.037 | 5.59 | 20.03 | 0.0060 | 0.2110 | 0.0307 | 0.4729 | 3.72 |
| M | 2.02 | 1.24 | 0.006 | 5175 | 3.037 | 5.59 | 20.12 | 0.0022 | 0.2110 | 0.0307 | 0.4731 | 3.72 |
| N | 2.03 | 1.24 | 0.006 | 5180 | 3.037 | 5.59 | 20.18 | 0.0043 | 0.2111 | 0.0307 | 0.4728 | 3.72 |
| O | 2.05 | 1.24 | 0.006 | 5190 | 3.037 | 5.58 | 20.33 | 0.0191 | 0.2113 | 0.0307 | 0.4726 | 3.72 |
| P | 1.97 | 1.25 | 0.006 | 5151 | 3.038 | 5.60 | 19.85 | 0.0021 | 0.2109 | 0.0306 | 0.4730 | 3.62 |
| Q | 1.98 | 1.25 | 0.006 | 5156 | 3.038 | 5.60 | 19.94 | 0.0013 | 0.2109 | 0.0307 | 0.4731 | 3.62 |
| R | 1.99 | 1.25 | 0.006 | 5162 | 3.038 | 5.60 | 20.03 | 0.0017 | 0.2110 | 0.0307 | 0.4732 | 3.62 |
| S | 2.01 | 1.25 | 0.006 | 5172 | 3.038 | 5.60 | 20.18 | 0.0080 | 0.2112 | 0.0307 | 0.4731 | 3.62 |
| T | 1.95 | 1.26 | 0.006 | 5142 | 3.038 | 5.62 | 19.86 | 0.0066 | 0.2109 | 0.0306 | 0.4737 | 3.52 |
| U | 1.96 | 1.26 | 0.006 | 5148 | 3.038 | 5.62 | 19.94 | 0.0054 | 0.2109 | 0.0307 | 0.4736 | 3.52 |
| V | 1.98 | 1.26 | 0.006 | 5159 | 3.039 | 5.62 | 20.09 | 0.0023 | 0.2111 | 0.0307 | 0.4732 | 3.52 |
| W | 1.96 | 1.27 | 0.006 | 5151 | 3.040 | 5.63 | 20.07 | 0.0127 | 0.2112 | 0.0307 | 0.4732 | 3.42 |

**Note.** The penultimate column represents the acoustic radius $\tau_0$ calculated using the asymptotic theory of stellar oscillations (Equations (38) and (39) from C. Aerts 2021 and Equation (5) from T. Wu & Y. Li 2016).

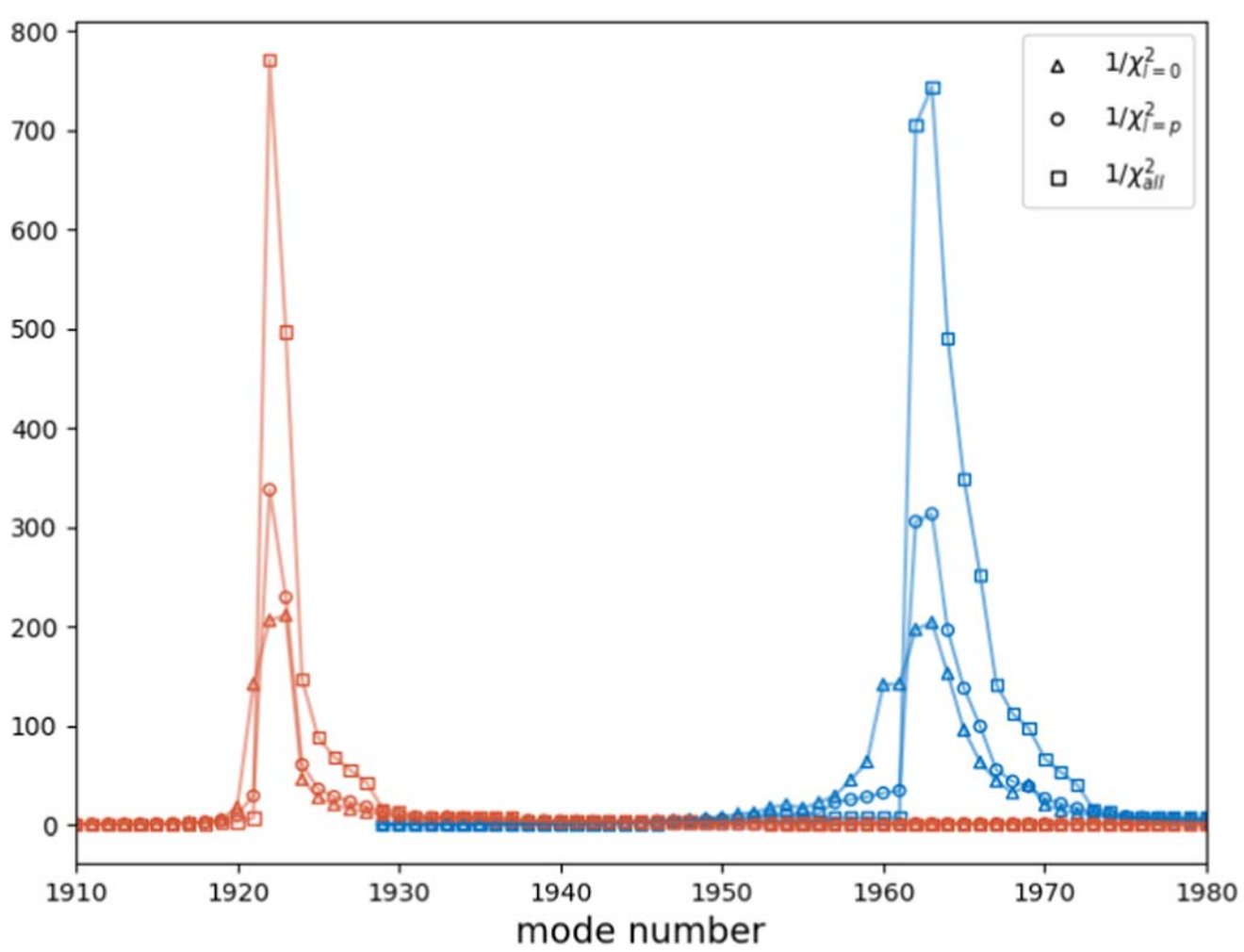

**Figure 5.** The evolution of the various $\chi^2$ parameters in Models E and Q throughout stellar evolution. Blue denotes Model E, while red represents Model Q. Triangles indicate $1/\chi^2_{l=0}$ as defined by Equation (8); circles denote $1/\chi^2_{l=p}$ as provided by $\chi^2_{l=p} = \frac{1}{N_{l=p}}\sum_{i=1}^{N_{l=p}}(\nu^{\rm obs}_{i(l=p)} - \nu^{\rm mod}_{i(l=p)})^2$ and $l = p$ denotes all $p$-modes, including $p$-dominated mixed modes; and squares represent $1/\chi^2_{\rm all}$ as defined by Equation (7).

surface. This reduction in temperature corresponds to a concurrent decrease in both pressure and density throughout the atmosphere. In MESA computations, once the temperature is determined based on the $T{-}\tau$ relationship, the pressure can be calculated using the equation of hydrostatic equilibrium. The resultant pressure comprises both gas pressure and radiation pressure. While gas pressure is derived from the equation of state, radiation pressure depends entirely on the fourth power of the temperature. Consequently, by adjusting the temperature, the ensuing variation in radiation pressure serves as a proxy for magnetic pressure, thereby simulating the influence of the magnetic field in the upper atmosphere. Figure 7 presents the distribution of both gas pressure and magnetic pressure for the optimal Model E. As pointed out by C. S. Rosenthal et al. (2002) and P. S. Cally (2007), $p$-mode oscillations will be reflected and transmitted at the interface where the gas pressure and magnetic pressure are roughly equal ($\beta = P_{\rm gas}/P_{\rm mag} \approx 1$). We employ this condition to infer the magnetic field strength within the stellar atmosphere. Based on Figure 7, we can deduce that the red giant KIC 9145955 has small-scale magnetic fields of approximately 65 G at a height of about 13,100 km in the photosphere.

On the other hand, the radius of KIC 9145955 is about 5.6 times that of the Sun and about 3 times that of HD 49385, so its atmosphere is correspondingly expanded by the same factor, which is consistent with the height of the small-scale magnetic field we have determined above.

Compared to the results of X. Zhang et al. (2018), the $\chi^2_{\rm all}$ value of our best-fit model (Model E) is reduced by about 50 times, which indicates that the frequency matching accuracy has been improved approximately 7 times. One of the main differences between our work and X. Zhang et al. (2018, 2022) is the treatment of the surface effect. Their works used the surface effect correction formula from I. M. Brandão et al. (2011), which is a power-law correction but considering the inertia of mixed modes and similar to the formula by H. Kjeldsen et al. (2008). However, our work do not apply any surface effect correction. Instead, we have altered the stellar surface structure by incorporating the effect of small-scale magnetic fields in the photosphere. Based on works of Y. Li et al. (2021) for the Sun and Y. Wang et al. (2024) for HD 49385, we have confirmed again that adding the effect of

**Table 2**
The Fundamental Parameters of KIC 9145955

| Variables | Observations | PH16 | Z18 | Z22 | Our Work |
|---|---|---|---|---|---|
| $T_{\rm eff}$ (K) | 4925 ± 91 | … | 5069 | 5032 ± 69 | 5167 ± 25 |
| $\log g$ | 3.04 ± 0.11 | … | 3.029 | 3.3031 ± 0.003 | 3.036 ± 0.004 |
| [Fe/H] | −0.32 ± 0.03 | … | … | −0.19 ± 0.13 | −0.48 |
| $M(M_\odot)$ | … | 1.196 | 1.24 | 1.23 | 1.23 ± 0.04 |
| $\alpha$ | … | 1.941 | 2.0 | 2.0 | 2.01 ± 0.05 |
| $f_{\rm OV}$ | … | 0.021 | 0 | 0.009 | 0 |
| $R(R_\odot)$ | … | 5.543 | 5.636 | 5.61 ± 0.004 | 5.57 ± 0.06 |
| $L(L_\odot)$ | … | 18.496 | 18.759 | 18.1 ± 0.8 | 19.85 ± 0.5 |
| Age (Gyr) | … | 3.912 | … | 4.61 ± 0.23 | 3.83 ± 0.5 |
| $M_{\rm He}(M_\odot)$ | … | … | 0.210 ± 0.002 | 0.2089 ± 0.0008 | 0.2108 ± 0.0005 |
| $R_{\rm He}(R_\odot)$ | … | … | 0.0307 ± 0.0002 | 0.03072 ± 0.00008 | 0.0306 ± 0.0001 |
| Height (km) | … | … | … | … | 13,100 |
| Strength (Gauss) | … | … | … | … | 65 |

**Note.** The observational data is sourced from F. Pérez Hernández et al. (2016). PH16 refers to the asteroseismic results by F. Pérez Hernández et al. (2016), while Z18 and Z20 represent the asteroseismic theoretical model results from X. Zhang et al. (2018) and X. Zhang et al. (2022), respectively.

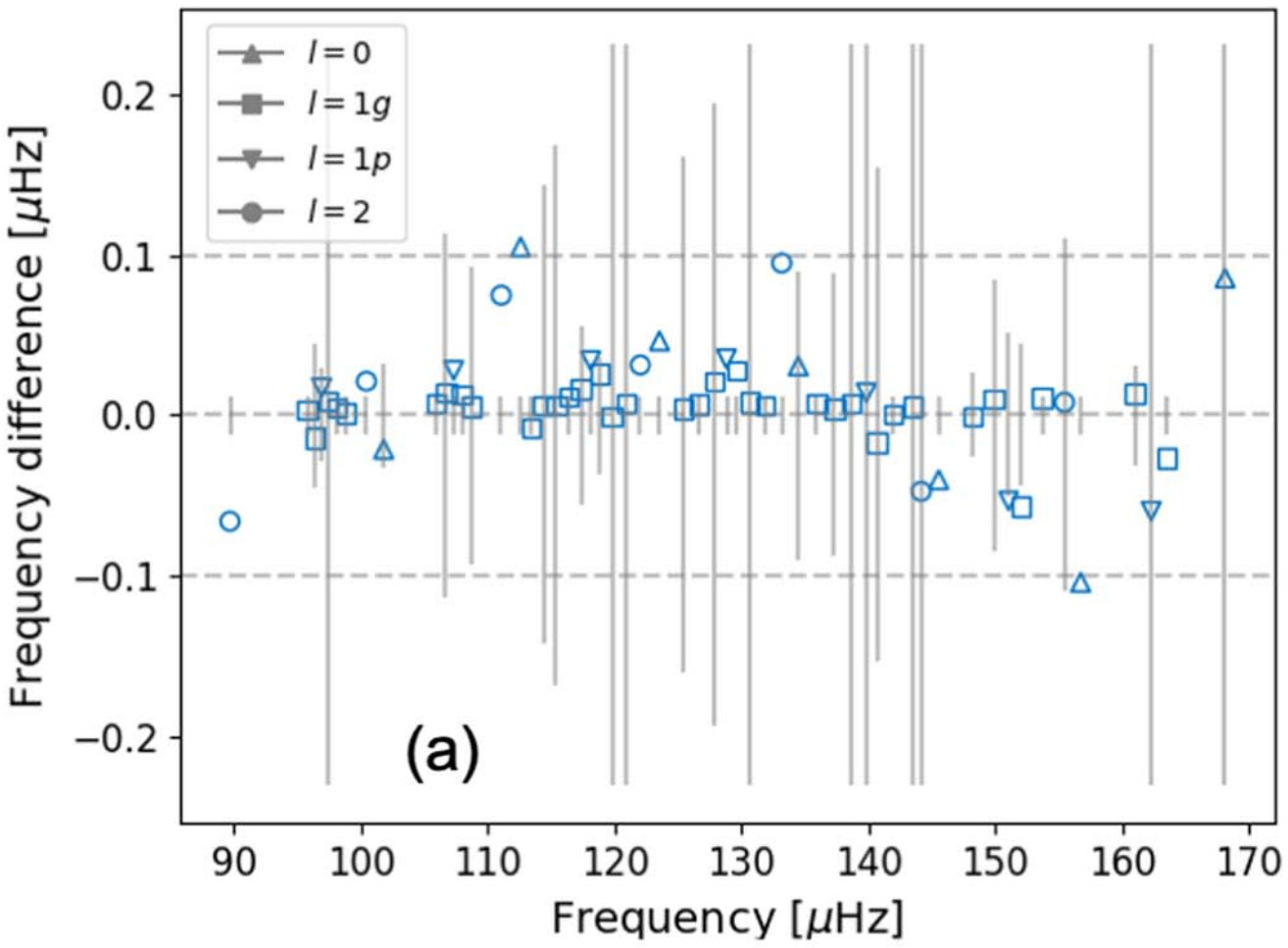

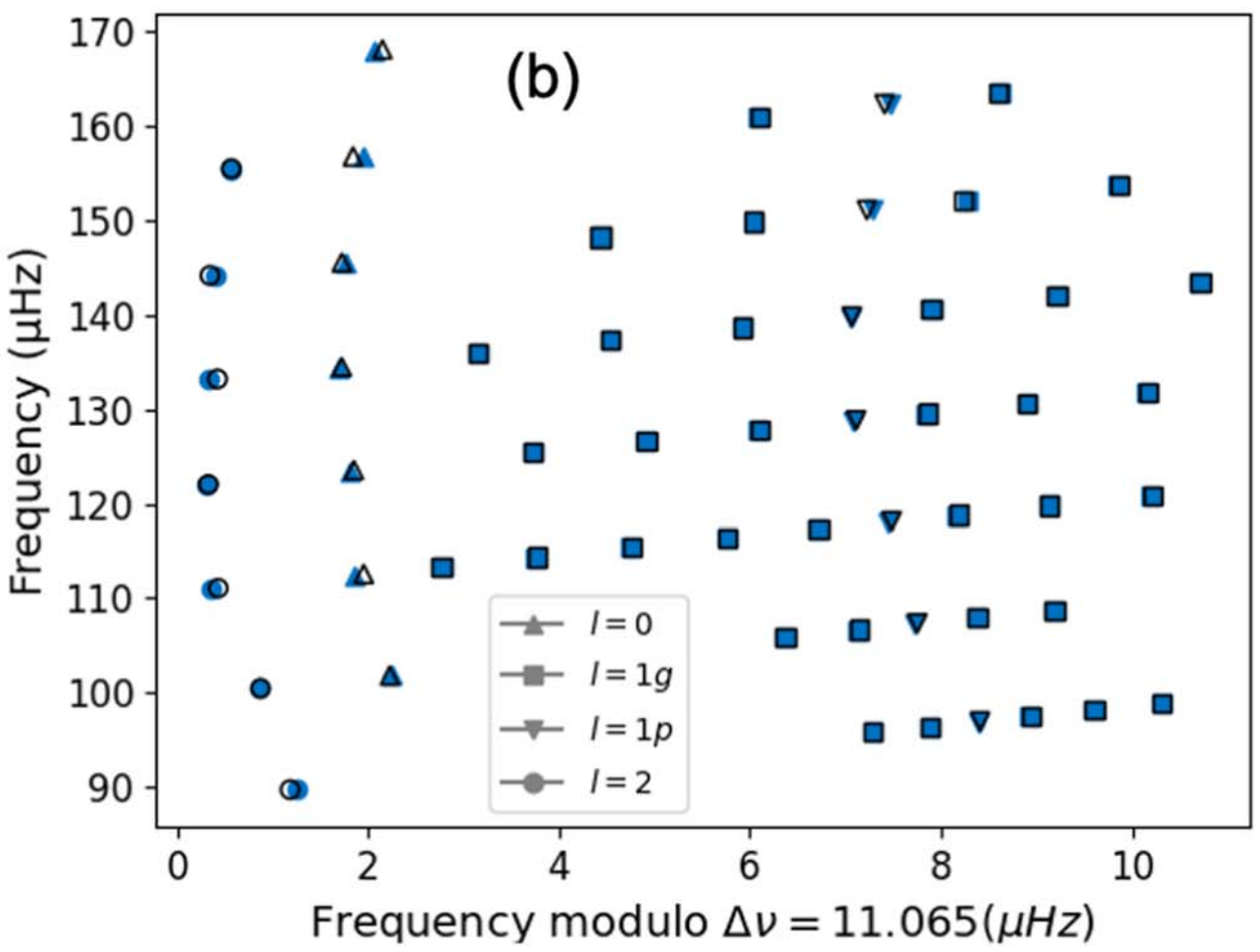

**Figure 6.** (a) Differences between the observed and theoretical frequencies of Model E. The gray vertical short lines represent the error bar of the observed frequencies. (b) Échelle diagram of the observed and calculated frequencies. The black dots represent the observational data, and the blue dots correspond to those of Model E.

small-scale magnetic fields in the photosphere of KIC 9145955 can improve the agreement between the frequencies of the best-fit theoretical model and the observed ones. However, numerous factors contribute to the surface effects, and given current technological limitations, we have only considered the influence of magnetic fields. Due to the omission of turbulent pressure, nonadiabaticity, and other physical effects, our study has certain limitations. Consequently, our estimates for the magnetic field should be regarded as rough upper limits.

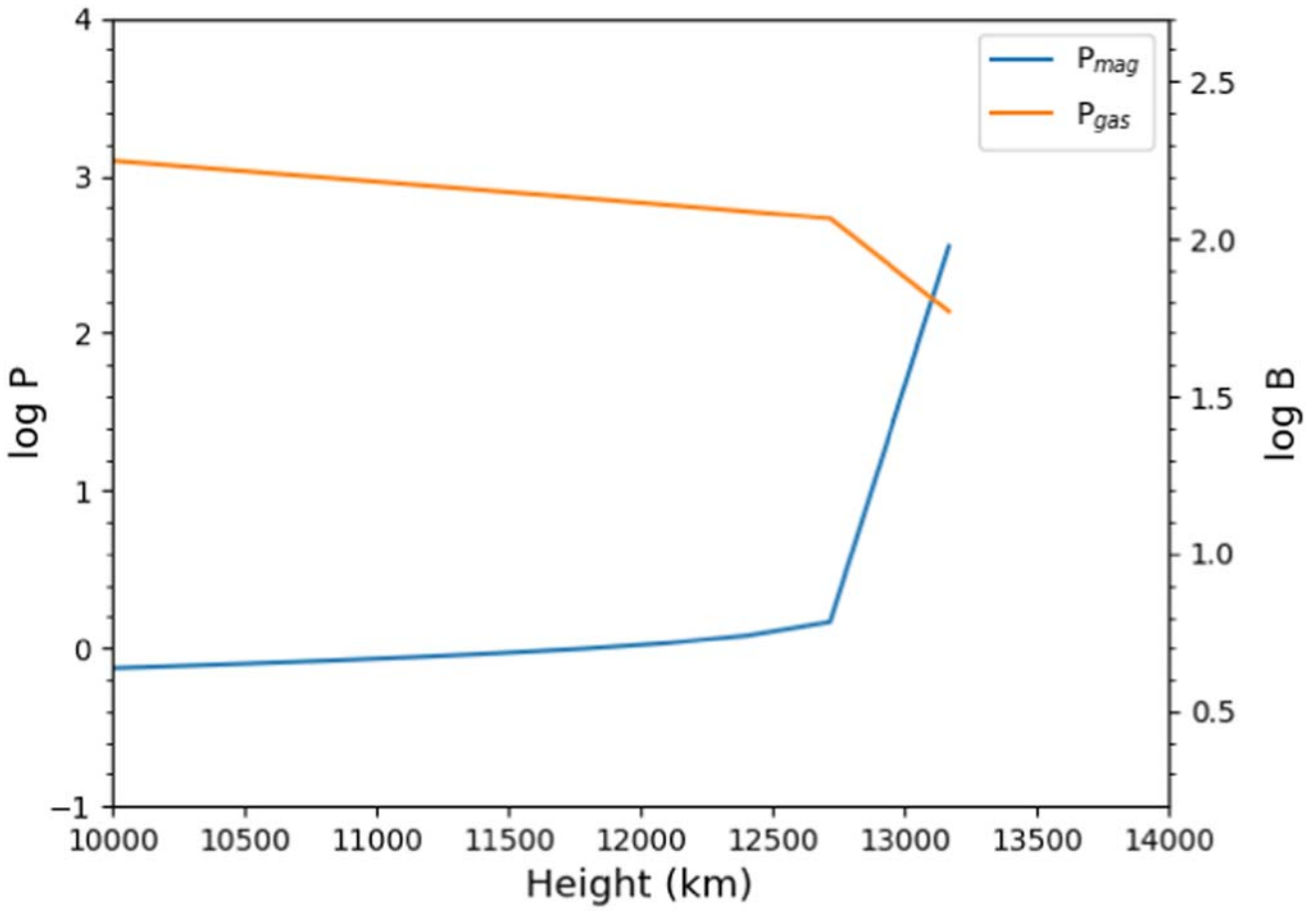

**Figure 7.** The variations of gas pressure and magnetic pressure with height in Model E. The zero height of the photosphere corresponds to $r = R$ (where $\tau = 2/3$).

Based on the data from Table 2, our best-fit model predicts a little higher effective temperature $T_{\rm eff}$ compared with the spectroscopic observations. This is likely due to the adoption of a lower [Fe/H] value than the observed one. A lower value of metallicity generally leads to a higher effective temperature. In comparison with the results of X. Zhang et al. (2018, 2022), the value of $\log g$ given by our best-fit model agrees better with the observations. As shown in Figure 4(a), our best candidate models exhibit strongly converged values of $\log g$. Since $\log g$ is determined by the mass and radius of a star, a reliable $\log g$ implies that the values of $M$ and $R$ derived from our best-fit model are also highly reliable. As for other stellar parameters like $\alpha$ and $L$, their values are similar to the results of X. Zhang et al. (2018, 2022). However, the age of KIC 9145955 obtained by us is slightly lower than their results, which can be attributed to the inclusion of convective core overshooting in their models. The presence of convective core overshooting slows down the stellar evolution process. Our estimated stellar age is comparable to the result of F. Pérez Hernández et al. (2016). Figure 4(b) also illustrates the fit quality dependence on the mass of the helium core. X. Zhang et al. (2022) showed that the observed frequencies of $l = 1g$ mixed modes are satisfactorily matched by stellar models with almost the same helium core mass. We confirm this conclusion by obtaining almost the same helium core mass in our work as in their result.

## 5. Conclusions

Red giants are huge stars with a compact helium core and an extended convective envelope. Solar-like oscillations are detected on tens of thousands of red giants, and this unique information provides a robust tool to explore the stellar structure and evolution. Recently, G. Li et al. (2022) found the evidence of magnetic fields ranging from 30 to 100 kG near the hydrogen-burning shell of three stars by analyzing the frequency shifts of $l = 1$ mixed. By incorporating the effect of magnetic fields into the solar atmospheric model, Y. Li et al. (2021) indicated the small-scale magnetic fields of 90 G at a height of 630 km in the solar photosphere. Red giants have the convective envelope, and are able to host small-scale magnetic fields, too. In the present paper, we choose KIC 9145955 as an example to explore the effect of the small-scale magnetic fields on its oscillation frequencies. KIC 9145955 is a well-studied star, with high quality observational data obtained from Kepler and well-determined basic parameters determined by carefully asteroseismic studies (F. Pérez Hernández et al. 2016; X. Zhang et al. 2018, 2022). By introducing a modified Eddington $T-\tau$ relation into the asteroseismic models to simulate the effect of small-scale magnetic fields in the photosphere, we carried out careful asteroseismic study for of KIC 9145955. The main conclusions are as follows:

1. The frequencies of our best-fit asteroseismic model match the frequencies of the observed $l = 0$, 1, and 2 oscillation modes with high precision. This demonstrates that incorporating small-scale magnetic fields in the atmosphere of KIC 9145955 helps address surface effect issues. Based on this, we determine a magnetic field with a strength of approximately 65 G, primarily concentrated at a height of about 13,100 km in the photosphere. Due to the omission of turbulent pressure, nonadiabaticity, and other physical effects, which provide additional support against gravity (C. S. Rosenthal et al. 1999; A. C. S. Jørgensen et al. 2021), the magnetic field strength derived from our model may approach the upper limit for KIC 9145955.
2. Based on perfect fitting of the observed frequencies, we can precisely determine the stellar parameters of KIC 9145955 as: $M = 1.23 \pm 0.04\, M_{\odot}$, $R = 5.57 \pm 0.06\, R_{\odot}$, $L = 19.85 \pm 0.5\, L_{\odot}$, Age $= 3.83 \pm 0.5$ Gyr, $M_{\rm He} = 0.2108 \pm 0.0005\, M_{\odot}$, and $R_{\rm He} = 0.0306 \pm 0.0001\, R_{\odot}$.

## Acknowledgments

This work has been cosponsored by National Natural Science Foundation of China (grant Nos. 12288102 and 12133011), National Key R&D Program of China (grant No. 2021YFA1600400/2021YFA1600402), the International Center of Supernovae at the Yunnan Key Laboratory (grant No. 202302AN360001), and the Natural Science Foundation of Yunnan Province (grant No. 202401CF070041). T.W. is grateful for the financial support from the B-type Strategic Priority Program of Chinese Academy of Sciences (grant No. XDB41000000), National Natural Science Foundation of China (grant No. 12273104), Yunnan Fundamental Research Projects (grant No. 202401AS070045), Youth Innovation Promotion Association of Chinese Academy of Sciences, and Yunnan Ten Thousand Talents Plan Young & Elite Talents Project. The authors also acknowledge the suggestions from X.-H. Chen and X.-Y. Zhang.

*Software:* MESA version 10398 (B. Paxton et al. 2011, 2013, 2015, 2018, 2019; A. S. Jermyn et al. 2023).

## Appendix
## Comparison of the Two Boundary Conditions

In this section, to verify the validity of using the Equation (4) as surface mechanical boundary condition, we provide an argument in a case "without" magnetic field that the two surface mechanical boundary conditions yield consistent results. We compare the differences between stellar pulsation models when using Equation (4) with the Eulerian approach and Equation (3) with the Lagrangian approach as pulsation boundary conditions. We performed calculations for the set of stellar parameters: $M = 1.23M_{\odot}$, $\alpha = 2.01$, and $Z = 0.006$. Figure 8 displays the Échelle diagram of the model frequencies corresponding to the minimum $\chi^2_{\rm all}$ value from the two evolutionary tracks. Both boundary conditions correspond to the same model (number 1985), and their calculated frequencies are virtually identical.

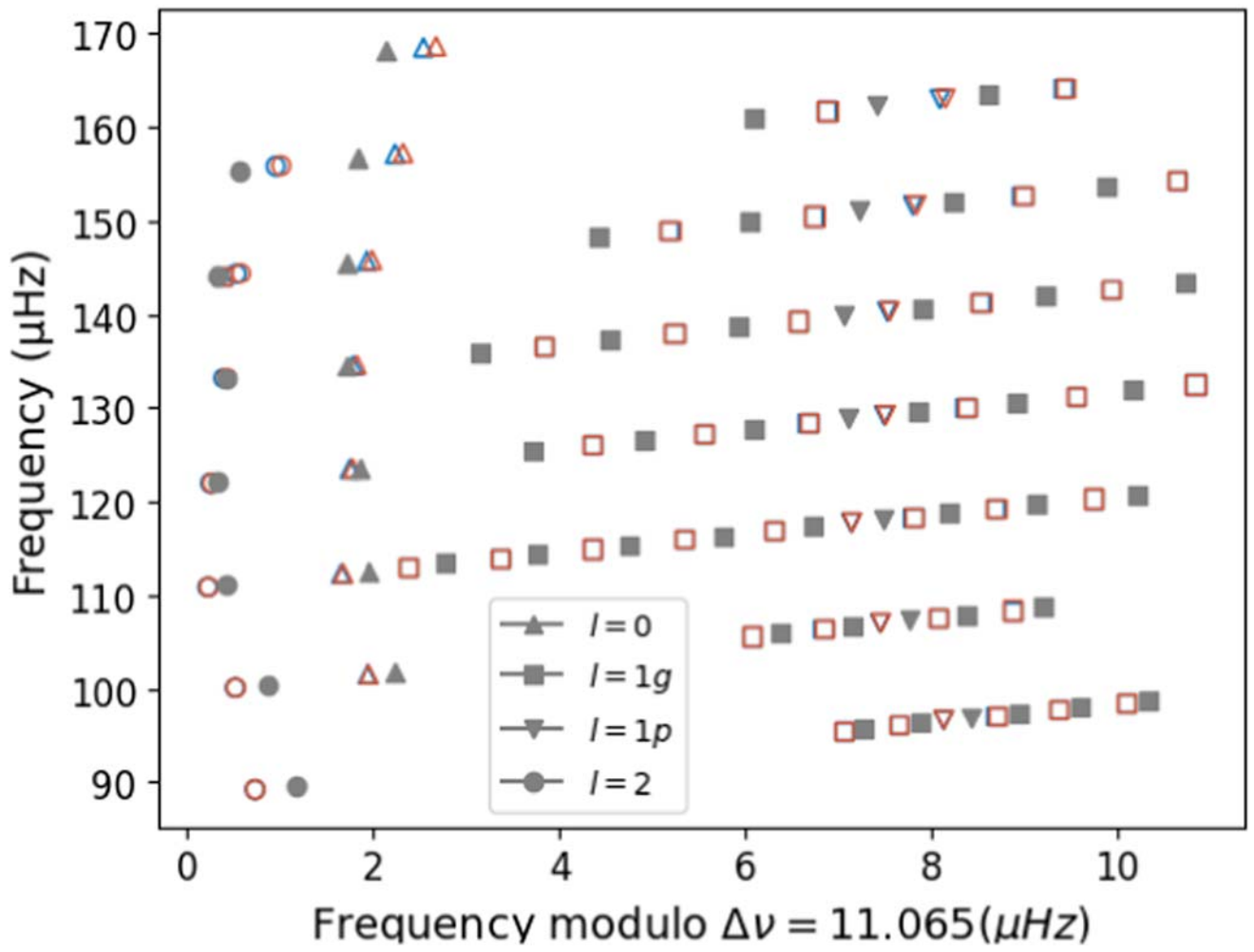

**Figure 8.** Échelle diagram of the model frequencies corresponding to the minimum $\chi^2_{\rm all}$ value (model number 1985) in the evolutionary tracks under the same set of stellar parameters for both boundary conditions. Blue and red hollow symbols, respectively, represent the case with no magnetic field using Equations (4) and (3) as boundary conditions. Gray solid symbols represent the observational results.

## ORCID iDs

Yuetong Wang https://orcid.org/0009-0005-2595-3750
Guifang Lin https://orcid.org/0000-0001-6725-8207
Yan Li https://orcid.org/0000-0002-1424-3164
Tao Wu https://orcid.org/0000-0001-6832-4325
Yaguang Li https://orcid.org/0000-0003-3020-4437